\documentclass[prl,reprint,longbibliography,superscriptaddress]{revtex4-1}

\usepackage{amsmath}
\usepackage{amssymb}
\usepackage{amsthm}

\usepackage{mathrsfs}
\usepackage{xcolor}
\usepackage{graphicx}

\usepackage{sourcesanspro}

\usepackage{phfqit}
\renewcommand{\DSym}{S}%
\renewcommand{\Dminz}[1][]{\DD_{\mathrm{min}}^{#1}}
\newcommand{\DKL}{\DD}
\DeclareMathOperator{\Var}{Var}

\newtheoremstyle{prlthm}%
  {0pt}%
  {0pt}%
  {\normalfont}%
  {\parindent}%
  {\itshape}%
  {:}%
  {0.8em}%
  {\thmname{#1\thmnumber{ #2}\thmnote{ #3}}}%

\newcounter{thm}

\newtheorem{maintheorem}[thm]{Theorem}

\usepackage{hyperref}%
\colorlet{linkcolor}{blue!50!black}
\hypersetup{bookmarksnumbered=false,bookmarksopen=false,bookmarksopenlevel=1,%
  breaklinks=true,pdfborder={0 0 0},colorlinks=true,%
  anchorcolor=linkcolor,citecolor=linkcolor,%
  filecolor=linkcolor,linkcolor=linkcolor,%
  menucolor=linkcolor,runcolor=linkcolor,%
  urlcolor=linkcolor}
\usepackage[all]{hypcap}%

\usepackage[nameinlink,capitalize]{cleveref}
\crefname{thm}{Theorem}{Theorems}

\renewcommand\paragraph[1]{%
  \par\emph{#1.---}\kern2pt\relax\ignorespaces}

\csname input\endcsname{bibolamazi_dup_aliases.tex}
 
\bibalias{Callen}{BookCallen_Thermo}
\bibalias{Lieb1999}{Lieb1999_secondlaw}
\bibalias{Nielsen}{BookNielsenChuang2000}
\bibalias{Esposito2009}{Esposito2009RMP_nonequilibrium}
\bibalias{Sagawa2012}{Sagawa2012LQC_secondlawlike}
\bibalias{Seifert2012}{Seifert2012RPP_stochastic}
\bibalias{Parrondo2015}{Parrondo2015NPhys_thermodynamics}
\bibalias{Bhatia}{BookBhatiaMatrixAnalysis1997}
\bibalias{Marshall}{BookMarshall2010Inequalities}
\bibalias{Blackwell}{Blackwell1951_comparison}
\bibalias{Ruch}{Ruch1978JCP}
\bibalias{Dupuis2012}{Dupuis2013_DH}
\bibalias{Horodecki2013}{Horodecki2013_ThermoMaj}
\bibalias{Brandao2015}{Brandao2015PNAS_secondlaws}
\bibalias{Gour}{Gour2015PR_resource}
\bibalias{Weilenmann2016}{Weilenmann2016PRL_axiomatic}
\bibalias{Gour2017}{Gour2017arXiv_entropic}
\bibalias{Faist2018}{Faist2018PRX_workcost}
\bibalias{Sagawa}{Sagawa-CMP-inprep}
\bibalias{Han2003}{BookHan_InfSpecMethods}
\bibalias{Han2000}{Han2000IEEETIT_hypothesis}
\bibalias{Nagaoka2007}{Nagaoka2007IEEETIT_hypothesis}
\bibalias{Datta_Renner}{Datta2009IEEE_InfSpec}
\bibalias{Datta2009}{Datta2009IEEE_minmax}
\bibalias{Bowen_Datta}{Bowen2006ISIT_beyondiid}
\bibalias{Bowen_Datta2}{Bowen2006arXiv_arbitrary}
\bibalias{Schoenmakers2007}{Schoenmakers2007ISIT_Renyi}
\bibalias{Renner2005}{PhDRenner2005_SQKD}
\bibalias{Matsumoto2010}{Matsumoto2010arXiv_reverse}
\bibalias{Jiao2017}{Jiao2018JMP_convertibility}
\bibalias{Weilenmann2017}{PhDMirjam2017}
\bibalias{Weilenmann2018}{Weilenmann2018arXiv_smooth}
\bibalias{Lieb2013}{Lieb2013_entropy_noneq}
\bibalias{Cover_Thomas}{BookCoverThomas2006_InfTheory}
\bibalias{Israel1979}{BookIsrael_ConvexityTheoryLatticeGases}
\bibalias{Ruelle1999}{BookRuelle_StatMechRigorous}
\bibalias{Bratteli1979}{BookBratteliRobinson_OpAlgQStatMech1}
\bibalias{Bratteli1981}{BookBratteliRobinson_OpAlgQStatMech2}
\bibalias{Tasaki2018}{Tasaki2018JSP_equivalence}
\bibalias{Alessio2016}{DAlessio2016AP_chaos}
\bibalias{Hiai_Petz}{Hiai1991CMP_proper}
\bibalias{Ogawa2000}{Ogawa2000IEEETIT_Stein}
\bibalias{Brandao2010}{Brandao2010CMP_Stein}
\bibalias{Bjelakovic2004}{Bjelakovic2004CMP_ergodic}
\bibalias{Bjelakovic2012}{Bjelakovic2003arXiv_revisted}
\bibalias{Watrous2009}{Watrous2009_sdps}
\bibalias{Bjelakovic2002}{Bjelakovic2004IM_lattice}
\bibalias{Bjelakovic2003}{Bjelakovic2003arXiv_compression}
\bibalias{Ogata2013}{Ogata2013LMP_shannonmcmillan}

\begin{document}

\title{Macroscopic Thermodynamic Reversibility in Quantum Many-Body Systems}

\author{Philippe Faist}
\affiliation{Institute for Quantum Information and Matter,
California Institute of Technology, Pasadena, CA 91125, USA}
\affiliation{Institute for Theoretical Physics, ETH Zurich, 8093 Switzerland}

\author{Takahiro Sagawa}
\affiliation{Department of Applied Physics, The University of Tokyo,
Tokyo 113-8656, Japan}

\author{Kohtaro Kato}
\affiliation{Institute for Quantum Information and Matter,
California Institute of Technology, Pasadena, CA 91125, USA}

\author{Hiroshi Nagaoka}
\affiliation{The University of Electro-Communications, 
Tokyo, 182-8585, Japan}

\author{Fernando G. S. L. Brand\~ao}
\affiliation{Institute for Quantum Information and Matter,
California Institute of Technology, Pasadena, CA 91125, USA}
\affiliation{Google Inc., Venice, CA 90291, USA}

\date{\today}

\begin{abstract}
  The resource theory of thermal operations, an established model for
  small-scale thermodynamics, provides an extension of equilibrium
  thermodynamics to nonequilibrium situations.  On a lattice of any dimension
  with any translation-invariant local Hamiltonian, we identify a large set of
  translation-invariant states that can be reversibly converted to and from the
  thermal state with thermal operations and a small amount of coherence.  These
  are the spatially ergodic states, i.e., states that have sharp statistics for
  any translation-invariant observable, and mixtures of such states with the
  same thermodynamic potential.  As an intermediate result, we show for a
  general state that if the min- and the max-relative entropy to the thermal
  state coincide approximately, this implies the approximately reversible
  interconvertibility to and from the thermal state with thermal operations and
  a small source of coherence.
  Our results provide a strong link between the abstract resource theory of
  thermodynamics and more realistic physical systems, as we achieve a robust and
  operational characterization of the emergence of a thermodynamic potential in
  translation-invariant lattice systems.
\end{abstract}

\maketitle

\paragraph{Introduction}%
The quantum information approach to thermodynamics has allowed thermodynamic
concepts, such as work, to be successfully extended into regimes of small-scale
systems that store and process quantum information~\cite{Goold2016JPA_review}.
Notably, formulating thermodynamics as a resource
theory~\cite{Janzing2000_cost,%
  Horodecki2003PRA_NoisyOps,%
  Brandao2013_resource,%
  Chitambar2018arXiv_resource} allows for a precise characterization of the
resources that are required in single-instance state transformations, for
instance thermodynamic work~\cite{Dahlsten2011NJP_inadequacy,%
  Aberg2013_worklike,Horodecki2013_ThermoMaj} and quantum
coherence~\cite{Lostaglio2015NC_beyond,Korzekwa2016NJP_extraction,%
  Gour2017arXiv_entropic,Marvian2018arXiv_distillation}.  This is done by
establishing a set of natural rules such as energy conservation, characterizing
which possible evolutions a quantum state can undergo under these rules, and
studying which external resources allows the system to undergo otherwise
forbidden state transformations.
A simple such framework is the resource theory of \emph{thermal operations},
where one allows any energy-conserving unitary interaction with a heat bath at a
fixed background temperature~\cite{Brandao2013_resource,%
  Horodecki2013_ThermoMaj,Brandao2015PNAS_secondlaws}, and can be extended to
more general types of reservoirs~\cite{YungerHalpern2016PRE_beyond,%
  Guryanova2016NatComm_multiple,%
  YungerHalpern2016NatComm_NATSandNATO,%
  HindsMingo2018arXiv_multiple}.  This approach has strong connections with
information-theoretic entropy measures and quantum Shannon
theory~\cite{Faist2015NatComm,Chubb2017arXiv_beyond}.  More generally,
information-theoretic approaches have provided new descriptions of
nonequilibrium states and dynamics in statistical mechanics and thermodynamics,
both in the classical and quantum regimes~\cite{Esposito2009,%
  Sagawa2012,Seifert2012,Parrondo2015,Tajima2016arXiv_large}.
The resource theory connects to standard macroscopic thermodynamics in several
ways.  This approach is equivalent~\cite{Weilenmann2016PRL_axiomatic,%
  PhDMirjam2017,Weilenmann2018arXiv_smooth} to an established abstract and
axiomatic formulation of thermodynamics by Lieb and
Yngvason~\cite{Lieb1999_secondlaw,Lieb2004_guide_secondlaw,%
  Lieb2013_entropy_noneq,Lieb2014PRSA_meter}.  Second, one recovers the usual
laws of thermodynamics in regimes of many identically and independently
distributed (i.i.d.\@) copies of a state, such as for an ideal gas, or if the
states considered are quantum statistical
ensembles~\cite{Brandao2013_resource,Horodecki2013_ThermoMaj,%
  Matsumoto2010arXiv_reverse,Jiao2018JMP_convertibility,%
  Faist2018PRX_workcost}.

The resource theory of thermodynamics extends equilibrium thermodynamics to
non-equilibrium situations.
In standard macroscopic thermodynamics, a system is defined to be in
thermodynamic equilibrium if it no longer presents macroscopic changes or
currents, and if it has lost memory of its initial, possibly non-equilibrium
state~\cite{BookCallen_Thermo}.  The purpose of this definition is to ensure
that the thermodynamic behavior of the system is entirely specified by a
\emph{thermodynamic potential}: The optimal work required to transform one
equilibrium state into another by a reversible thermodynamic process is given by
the difference of the potentials for the initial and final states, and does not
depend on any further details of the process.
In the resource theory, this can be verified directly: Is the amount of work
required to transform a state $A$ into a state $B$ equal to the amount of work
that can be extracted in the reverse process?  If so, the resource theory is
said to be \emph{reversible}.  Crucially, reversibility of a resource
theory---i.e., the emergence of a thermodynamic potential---can happen for
states that are not necessarily in thermodynamic equilibrium, as we show in this
paper.

A natural question is whether the notion of resource-theoretic reversibility can
be leveraged to show the emergence of a thermodynamic potential for new classes
of states that are physically relevant, such as interacting particles on a
lattice, which go beyond idealized macroscopic settings such as i.i.d.\@ states.

Here, we show that on a translation-invariant lattice of any spatial dimension
with a local Hamiltonian, all ergodic states---i.e., states for which
macroscopic quantities have sharply peaked statistics---can be reversibly
converted to and from the thermal state.  Furthermore, mixtures of ergodic
states with the same thermodynamic potential also have this property.  This
ensures the emergence of a thermodynamic potential for this class of states even
if some of these states are far out of equilibrium.

In the following, we first introduce the resource theory of thermodynamics and
show that for general states, an equipartition property implies the emergence of
a thermodynamic potential.  We then consider translation-invariant lattices and
explain our main result illustrated with an example of a 1-D Ising spin chain,
before concluding with a discussion.

\paragraph{Resource theory of thermal operations}%
In this resource theory, one is allowed to (i)~bring in any ancilla systems in
their thermal state, (ii)~to carry out any energy-conserving unitaries, and
(iii)~to trace out any systems.
We may then quantify the amount of work required to transform $\rho$ into
another state $\rho'$ by including an explicit \emph{battery system},
initialized in a pure energy eigenstate $\ket{E}$ and which we require to
transition into another energy eigenstate $\ket{E'}$ at the end of the process.
That is, if the transformation $\rho\otimes\proj{E} \to \rho'\otimes\proj{E'}$
is possible with the operations (i)--(iii), then we define this process as
consuming $E-E'$ work~\cite{Horodecki2013_ThermoMaj,%
  Skrzypczyk2014NComm_individual,Faist2018PRX_workcost} (negative work
consumption corresponds to work extraction).

\begin{figure}
  \centering
  \includegraphics{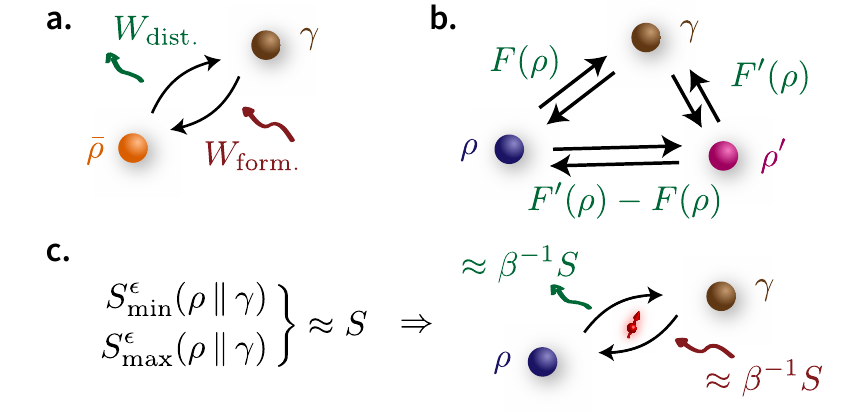}
  \caption{A thermodynamic potential emerges when the underlying resource theory
    is reversible.  \textbf{a.}~For a state $\bar\rho$ that is block-diagonal in
    energy, the work that can be extracted is given by the min-relative entropy
    $W_{\mathrm{dist.}} = \beta^{-1}\Dminz[\epsilon]{\bar\rho}{\gamma}$, leaving
    the system in the thermal state $\gamma=\ee^{-\beta H}/\tr(\ee^{-\beta H})$.
    Conversely, the work required to prepare $\bar\rho$ from the thermal state
    is $W_{\mathrm{form.}} = \beta^{-1}\Dmax[\epsilon]{\bar\rho}{\gamma}$.
    \textbf{b.}~Suppose a state $\rho$ (respectively $\rho'$) can be reversibly
    converted to and from the thermal state with work $F(\rho)-F(\gamma)$
    (respectively $F(\rho')-F(\gamma)$).  Then $\rho$ and $\rho'$ can be
    reversely interconverted.  In this case the resource theory is said to be
    \emph{reversible}, and the thermodynamic potential $F(\rho)$ fully
    characterizes the work required for state transformations.  \textbf{c.}~As
    an intermediate result, we show that if the min- and the max-relative
    entropies of any arbitrary quantum state $\rho$ coincide approximately, then
    coherences in the state are suppressed, making it nearly block-diagonal in
    energy.  The state is then approximately reversibly convertible to and from
    the thermal state with thermal operations and a small source of coherence.}
  \label{fig:ReversibleInterConversionStates}
\end{figure}

We refer to the class of states which are block-diagonal in the energy
eigenspaces as \emph{semiclassical states}.  For these states, transformations
under thermal operations are fully characterized by
\emph{thermo-majorization}~\cite{Horodecki2013_ThermoMaj}, a generalized notion
of matrix majorization~\cite{Ruch1978JCP,BookBhatiaMatrixAnalysis1997,%
  BookMarshall2010Inequalities}.
Let's consider two natural tasks associated with a semiclassical state $\rho$:
\emph{state formation} and \emph{work distillation}
(\cref{fig:ReversibleInterConversionStates}\textbf{a}).
State formation consists in preparing the state $\rho$ starting from the thermal
state of the system, $\gamma=\ee^{-\beta H}/\tr(\ee^{-\beta H})$.  The optimal
amount of work that needs to be invested, if we allow an inaccuracy $\epsilon>0$
in the final state and if $\rho$ is semiclassical,
is~\cite{Aberg2013_worklike,Horodecki2013_ThermoMaj}
\begin{align}
  W_{\mathrm{formation}}(\rho) = \beta^{-1}\Dmax[\epsilon]{\rho}{\gamma}\ ,
  \label{eq:Wformation-Dmax}
\end{align}
with the \emph{max-relative entropy} defined as
$\Dmax[\epsilon]{\rho}{\sigma}
= \min_{\tilde\rho\approx_\epsilon\rho}
\ln\, \norm[\big]{ \sigma^{-1/2}\,\tilde\rho\,\sigma^{-1/2} }_{\infty}$
with the optimization ranging over all states $\tilde\rho$ that are
$\epsilon$-close to $\rho$ in trace distance~\cite{Datta2009IEEE_minmax}.  
On the other hand, \emph{work distillation} consists in extracting as much work
as possible from a given state $\rho$, resulting in the thermal state $\gamma$
on the system.  The optimal amount of work that can be extracted from a
semiclassical state $\rho$ is~\cite{Aberg2013_worklike,Horodecki2013_ThermoMaj}
\begin{align}
  W_{\mathrm{distillable}}(\rho) =
  \beta^{-1}\Dminz[\epsilon]{\rho}{\gamma}\ ,
\end{align}
with the \emph{min-relative entropy} defined as
$\Dminz[\epsilon]{\rho}{\sigma}
= \max_{\tilde\rho\approx_\epsilon\rho} \left\{
  -\ln\, \tr\bigl(\Pi^{\tilde\rho}\,\sigma\bigr) \right\}$
where $\Pi^{\tilde\rho}$ is the projector onto the support of
$\tilde\rho$~\cite{Datta2009IEEE_minmax}.
The min- and max-relative entropies are special cases of the R\'enyi relative
entropies~\cite{Renyi1960_MeasOfEntrAndInf,%
  PhDTomamichel2012,BookTomamichel2016_Finite}.

There are no known necessary and sufficient conditions for transformations of
arbitrary states under thermal operations.  The reason is that thermal
operations 
cannot generate any coherent superposition of energy levels, underscoring the
role of time asymmetry in thermodynamics~\cite{%
  Aberg2014PRL_catalytic,%
  Marvian2014PRA_modes,%
  Marvian2014NC_extending,%
  Lostaglio2015NC_beyond,%
  Korzekwa2016NJP_extraction,%
  Gour2017arXiv_entropic,%
  Marvian2018arXiv_distillation,%
  Lostaglio2018arXiv_broadcast}.  It is thus necessary to account for coherence
as a separate resource that enable operations that cannot be performed with
thermal operations alone~\cite{Baumgratz2014PRL_coherence,%
  Winter2016PRL_coherence,%
  Dana2017PRA_beyond,%
  Diaz2018Qu_reusing,%
  Popescu2018arXiv_applications}.

We resort to a very rudimentary way of accounting for coherence.  We allow a
system $C$ with a bounded range of energy, which can be prepared in any pure
state of our choosing and which we must dispose of in any state that is close to
a pure state.  This energy range is what we refer to as \emph{amount of
  coherence} when such a system is used in a thermodynamic process.  This crude
approach is sufficient for our purposes, since our protocols only require such a
system with an energy range that is negligibly small compared to the overall
work cost of the transformation, thus forbidding any noticeable embezzling of
work~\cite{Brandao2015PNAS_secondlaws}.

\paragraph{Emergence of a thermodynamic potential}%
A resource theory is
\emph{reversible} for a class of states if the optimal work cost of any
transition between two such states is equal to the optimal work extracted in the
corresponding reverse process.  This class of states then has a total order, and
we can assign a ``thermodynamic value'' to each state---this is the
thermodynamic potential.
A sufficient condition for reversibility is to check whether the work required
for state formation can fully be recovered in the reverse task of work
distillation~\cite{Horodecki2013_ThermoMaj}; any transformation between two such
states is then reversible
(\cref{fig:ReversibleInterConversionStates}\textbf{b}).
In well-behaved cases, such as in the i.i.d.\@
regime~\cite{Brandao2013_resource} or for statistical
ensembles~\cite{Weilenmann2018arXiv_smooth}, the thermodynamic potential is
given by the Kullback-Leibler divergence or Umegaki relative entropy
$\DKL{\rho}{\gamma}$, defined as
\begin{align}
  \DKL{\rho}{\sigma} = \tr\bigl( \rho\bigl(\ln\rho - \ln\sigma\bigr) \bigr)\ .
  \label{eq:KL-div-defn}
\end{align}

\paragraph{Equipartition implies reversibility with thermal operations}%
We first present an intermediate result: If the min- and max-relative entropies
coincide approximately, a condition which can be interpreted as a form of
equipartition, then the state can approximately be reversibly converted to and
from the thermal state (\cref{fig:ReversibleInterConversionStates}\textbf{c}).
Our physical explanations are complemented by a fully rigorous proof that will
be published elsewhere~\cite{Sagawa-CMP-inprep}.

\begin{maintheorem}
  \label{mainthm:equipartition-implies-reversibility-by-TO}
  For any $\rho$ and for $\epsilon>0$, suppose that
  \begin{align*}
    \Dminz[\epsilon]{\rho}{\gamma} &\geqslant S-\Delta\ ;
    &
    \Dmax[\epsilon]{\rho}{\gamma} &\leqslant S+\Delta\ ,
  \end{align*}
  for some $S\in\mathbb{R}, \Delta>0$.  Then $\rho$ can be approximately
  converted to and from the thermal state at a work cost (resp.\@ work yield) of
  approximately $\beta^{-1}[S+O(\Delta)]$ (resp.\@ $\beta^{-1}[S-O(\Delta)]$),
  with an amount of coherence of approximately $O(\Delta)$, and with arbitrarily
  good precision as $\epsilon\to0$.
\end{maintheorem}

For a system of $n$ particles, if we have $\Delta/n \to 0$ as $n\to\infty$, then
the extractable work per system and the work of formation per system both
converge to $s_\infty(\rho) := \lim_{n\to\infty} S/n$, and the amount of
coherence used per copy goes to zero.  In this case $s_\infty(\rho)$ becomes the
thermodynamic potential in the thermodynamic limit $n\to\infty$.

To prove \cref{mainthm:equipartition-implies-reversibility-by-TO}, we first show
that a state $\rho$ for which the min-entropy and the max-entropy differ by at
most $O(\Delta)$ have off-diagonal elements $\matrixel{E_k}{\rho}{E_{k'}}$ that
are exponentially suppressed in $\beta\abs{E_k-E_{k'}}$ if
$\beta\abs{E_k-E_{k'}}\gtrsim O(\Delta)$.  In this sense, such a state may not
harbor a large amount of coherence.
\Cref{mainthm:equipartition-implies-reversibility-by-TO} is then proven by
exhibiting protocols for work distillation and state formation with the claimed
properties.
For both protocols, we first replace the Hamiltonian by one where the energy
levels are integer multiples of some elementary spacing $O(\Delta)$, which can
be done by investing an amount of coherence of order $O(\Delta)$.
The work distillation protocol is then executed as follows.  One dephases $\rho$
in the new energy basis.  Then we apply the known protocol for work extraction
of semiclassical states.  Because $\rho$ has little coherence, the work that was
wasted by the dephasing is small and the min-entropy does not change by much, so
we can still recover $S-O(\Delta)$ work.
For the second protocol, we use the notion of an \emph{internal reference
  frame}: The state $\rho$ is equivalently described by a completely incoherent
state $\tilde{\rho} = \mathcal{D}[\rho\otimes\eta]$, where $\eta$ is a special
state called a \emph{reference frame}, and where $\mathcal{D}[\cdot]$ is the
joint dephasing operation on the system and the reference
frame~\cite{Bartlett2006IJQI_dialogue,Bartlett2007_refframes}.  Because $\rho$
has only little coherence, a small reference frame $\eta$ suffices to achieve an
accurate description of $\rho$.  Our protocol consists in first preparing the
incoherent state $\tilde{\rho}$ using the known protocol for semiclassical
states, and then ``shifting'' the coherence from $\eta$ to $\rho$, a process
known as ``externalizing'' the reference frame~\cite{Bartlett2007_refframes}.

\paragraph{Ergodic states on a lattice}%
We now consider a $d$-dimensional square lattice with a local Hamiltonian that
is translation-invariant:
\begin{align}
  H = \sum_{\boldsymbol{z}\in\mathbb{Z}^d} h_{\boldsymbol{z}}\ ,
\end{align}
where each term $h_{\boldsymbol{z}}$ is a lattice-translated version of a term
$h_{\boldsymbol{0}}$ that acts on a constant number of sites neighboring the
origin.  Each site is a quantum system of some finite dimension.  
Our calculations will be performed for finite lattice sizes, where the total
number of sites is denoted by $n$.  For finite $n$, the Hamiltonian is truncated
at the boundary by ignoring any terms that have support outside of the finite
region considered.

\begin{figure}
  \centering
  \includegraphics{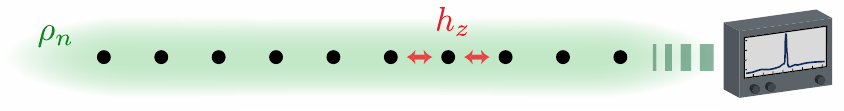}
  \caption{Ergodic state on a lattice.  An ergodic state is one that is
    translation-invariant and that produces sharp statistics for any
    translation-invariant observable.  Our main result is to show that any two
    ergodic states can be reversibly interconverted with thermal operations and
    a sublinear amount of coherence, with the reversible work cost deriving from
    a thermodynamic potential given by the Kullback-Leibler divergence.
    Furthermore, a translation-invariant state has a thermodynamic potential if
    and only if it is a mixture of ergodic states of equal potential, providing
    a robust and operational understanding of the emergence of a thermodynamic
    potential in lattice systems.}
  \label{fig:LatticeErgodicState}
\end{figure}

In statistical mechanics, thermodynamic behavior is often captured in the notion
of \emph{ergodicity} (\cref{fig:LatticeErgodicState}).
Ergodic states are defined on the infinite lattice in two equivalent
ways~\cite{BookRuelle_StatMechRigorous,%
  BookBratteliRobinson_OpAlgQStatMech1,%
  BookBratteliRobinson_OpAlgQStatMech2,%
  Bjelakovic2004CMP_ergodic,Bjelakovic2004IM_lattice}.  First, they are exactly
those states that self-average over space translations. 
I.e., an ergodic state $\rho$ satisfies the following property: For any local
observable $a_{\boldsymbol{0}}$,
we have $\Var_\rho\bigl(\frac1n\sum a_{\boldsymbol{z}}\bigr) \to 0$ as
$n\to\infty$.
Equivalently, ergodic states are the extremal points of the set of states that
are translation-invariant on the infinite lattice.  Consequently, any
translation-invariant state can be written as a mixture of ergodic states.

Ergodic states are the natural quantum analogue of classically ergodic
probability distributions~\cite{BookCoverThomas2006_InfTheory,%
  BookIsrael_ConvexityTheoryLatticeGases} for spatial translations instead of
time evolution.
Examples of ergodic states include Gibbs states of a local Hamiltonian at
sufficiently high temperature, where the correlation functions of local
observables decay exponentially in space (see for example Ref.~\cite{Tasaki2018}
and references therein).  Also, any i.i.d.\@ state is ergodic, being the Gibbs
state of a noninteracting Hamiltonian.  In contrast, a mixed state of
macroscopically different sectors (e.g., different magnetization sectors in a
symmetry-broken phase) is not ergodic, as spatial fluctuations do not vanish.

\paragraph{Ergodicity and reversibility under thermal operations}%
Our main contribution is to prove that on a lattice of any dimension with a
translation-invariant local Hamiltonian, all ergodic states fall into the
setting of \cref{mainthm:equipartition-implies-reversibility-by-TO} and are
thus reversibly interconvertible:

\begin{maintheorem}
  \label{mainthm:ergodic-state-has-thermo-potential}
  In the thermodynamic limit $n\to\infty$, any two ergodic states can be
  reversibly converted into one another using thermal operations and a sublinear
  amount of coherence, and the corresponding reversible work cost rate is given
  by the thermodynamic potential
  \begin{align}
    s(\rho) = \lim_{n\to\infty} \frac1n \DKL{\rho_n}{\gamma_n}\ ,
    \label{eq:mainthm-ergodic-state-thermo-potential-KL-divergence}
  \end{align}
  where $\rho_n$ is the reduced state of $\rho$ on a finite sublattice of size
  $n$ and $\gamma_n = \ee^{-\beta H_n} / \tr(\ee^{-\beta H_n})$ is the Gibbs
  state with the truncated Hamiltonian $H_n$ on the sublattice.
\end{maintheorem}

The proof of \cref{mainthm:ergodic-state-has-thermo-potential} proceeds via the
\emph{hypothesis testing relative entropy}~\cite{Buscemi2010IEEETIT_capacity,%
  Brandao2011IEEETIT_oneshot,%
  Wang2012PRL_oneshot,%
  Matthews2014IEEETIT_blocklength,%
  Tomamichel2013_hierarchy,%
  Dupuis2013_DH}, which interpolates between the min- and max-relative
entropies~\cite{Dupuis2013_DH} and can be formulated as a semidefinite
program~\cite{Watrous2009_sdps}.  Inspired by the proof techniques of~\cite{%
  Bjelakovic2003arXiv_revisted,%
  Bjelakovic2003arXiv_compression,%
  Bjelakovic2004CMP_ergodic,%
  Bjelakovic2004IM_lattice,%
  Ogata2013LMP_shannonmcmillan}, we construct a quantum relative typical
projector for an ergodic state relative to a Gibbs state associated with a
truncated local Hamiltonian.  This allows us to prove a generalized version of
Stein's lemma for hypothesis testing~\cite{%
  Hiai1991CMP_proper,%
  Ogawa2000IEEETIT_Stein,%
  Bjelakovic2003arXiv_revisted,%
  Bjelakovic2004CMP_ergodic,%
  Brandao2010CMP_Stein} from which it follows that the min- and max-relative
entropies must coincide up to sublinear terms in $n$, and where the limiting
value converges to $s(\rho)$.
We are then in the setting of
\cref{mainthm:equipartition-implies-reversibility-by-TO}: Any ergodic state can
be reversibly converted to and from the thermal state with the reversible work
deriving from the thermodynamic potential $s(\rho)$.
A rigorous proof will be published elsewhere~\cite{Sagawa-CMP-inprep}.

\paragraph{Translation-invariant states and reversibility}
We can further ask, is there a larger class of translation-invariant states on a
lattice that can be reversibly converted to and from the thermal state?  We
provide an answer to this question as follows:

\begin{maintheorem}
  \label{mainthm:thermo-potential-iff-mix-ergodic-states-eq-potential}
  A translation-invariant state $\rho$ that is a mixture of a finite number of
  ergodic states is reversibly convertible to and from the thermal state if and
  only if all ergodic states in the mixture are of equal potential, i.e.,
  $\rho=\sum p_k \rho^{(k)}$ with $s(\rho^{(1)}) = s(\rho^{(2)}) = \cdots$.
\end{maintheorem}

To prove the above theorem, we note the following property of the min- and
max-relative entropy for a mixture $\rho=\sum p_k \rho^{(k)}$:
\begin{subequations}
  \begin{align}
    \Dminz[\epsilon]{\rho_n}{\gamma_n}
    &\approx \min_k \Dminz[\epsilon']{\rho_n^{(k)}}{\gamma_n}\ ;
    \\
    \Dmax[\epsilon]{\rho_n}{\gamma_n}
    &\approx \max_k \Dmax[\epsilon']{\rho_n^{(k)}}{\gamma_n}\ ,
  \end{align}
\end{subequations}
with the approximation holding up to terms that do not scale with $n$ and up to
an adjustment of the smoothing parameter $\epsilon$ that does not depend on $n$.
If all the $\rho^{(k)}$ in the decomposition have the same potential,
$S(\rho^{(1)}) = S(\rho^{(2)}) = \cdots$, then
$\Dminz[\epsilon]{\rho_n}{\gamma_n} \approx \Dmax[\epsilon]{\rho_n}{\gamma_n}$
with equality in the thermodynamic limit, and we can apply
\cref{mainthm:equipartition-implies-reversibility-by-TO}.  Conversely, if the
$\rho^{(k)}$ do not all have the same potential, then the min- and max-relative
entropies differ even in the thermodynamic limit.  This implies that $\rho$
cannot be reversibly convertible to and from the thermal state, because the min-
and max-relative entropies are monotones under thermal operations.

\paragraph{Example: 1D Ising spin chain}%
This toy example illustrates how a thermodynamic potential can emerge for states
that are not in thermodynamic equilibrium.
Consider a 1D chain of spin-1/2 particles with an Ising nearest-neighbor
(n.n.\@) coupling and an external field $h$:
\begin{align}
  H = -J\! \sum_{i, j\text{ n.n.}}\! \sigma_z^{i}\sigma_z^{j} + h\sum_i \sigma_z^{i}\ ,
\end{align}
where $\sigma_z = \proj\uparrow - \proj\downarrow$.  Since i.i.d.\@ states are
ergodic, our results imply that two pure states of the form
$\ket{\psi}^{\otimes n}, \ket{\psi'}^{\otimes n}$ can be converted into one
another with thermal operations and an asymptotically negligible source of
coherence at a reversible work cost of $F_\psi - F_{\psi'}$ per copy, where the
thermodynamic potential is $F_\psi = \beta^{-1} \lim_{n\to\infty} %
\DKL{\psi^{\otimes n}}{\gamma_n}/n$, which is the free energy per site up to an
unimportant additive constant.  The thermodynamic potential is well defined on
an operational level even for states $\psi^{\otimes n}$ that are not in
macroscopic equilibrium.  Consider for instance the state
$\ket\psi=\ket+=[\ket\uparrow+\ket\downarrow]/\sqrt{2}$.  For $h\gg J$, the
state $\psi^{\otimes n}$ presents macroscopic changes in the total spin along
the $X$ axis under time evolution according to $H$, but this does not prevent it
from being reversibly convertible to and from another state
$\ket{\psi'}^{\otimes n}$.

\paragraph{Discussion}%
Our results provide a direct link between the abstract theory of thermodynamics
at the small scale formulated in terms of a resource theory, and realistic
many-body systems that are commonly studied in statistical mechanics. 
In statistical mechanics, an ergodic state physically corresponds to a definite
macroscopic state; it describes a pure thermodynamic phase without phase
coexistence~\cite{Ruelle1999}.
We endow these ergodic states with a stronger notion of thermodynamic behavior:
The notion of reversibility associated with the resource theory---which extends
the concept in equilibrium thermodynamics to nonequilibrium
situations---is tightly related to the notion of ergodicity.
Furthermore, our analysis underscores how reversibility in the resource theory
does not imply equilibrium.  Indeed, spatially ergodic states, as considered here,
can evolve nontrivially in time as illustrated in the toy example
above.

Our rigorous proof~\cite{Sagawa-CMP-inprep} makes use of advanced
information-theoretic techniques, including the information
spectrum~\cite{BookHan_InfSpecMethods,%
  Han2000IEEETIT_hypothesis,%
  Nagaoka2007IEEETIT_hypothesis,%
  Datta2009IEEE_InfSpec,%
  Bowen2006ISIT_beyondiid,%
  Bowen2006arXiv_arbitrary,%
  Schoenmakers2007ISIT_Renyi}, hypothesis testing and quantum Stein's
lemma~\cite{Hiai1991CMP_proper,%
  Ogawa2000IEEETIT_Stein,%
  Tomamichel2013_hierarchy,%
  Dupuis2013_DH}, as well as quantum typical projectors~\cite{%
  BookWilde2013QIT,Bjelakovic2003arXiv_revisted,%
  Bjelakovic2004IM_lattice,Bjelakovic2004CMP_ergodic}.
Our results can be seen as an extension of the ergodic theorems of
Refs.~\cite{Bjelakovic2004CMP_ergodic,Bjelakovic2004IM_lattice}.
We also use Ref.~\cite{Lenci2005JSP_onephase} to show that if we consider the
reduced state of the infinite-dimensional Gibbs state instead of truncating the
Hamiltonian for finite sublattices, then our results persist for sufficiently
high temperatures where there is a unique KMS state.

Curiously, it is possible to construct toy situations in which the thermodynamic
potential is not given by the Kullback-Leibler
divergence~\cite{Sagawa-CMP-inprep}.
While this does not happen in the setting considered in the present paper, it
shows that the Kullback-Leibler divergence is not universally the correct
expression of the emergent thermodynamic potential as defined via
\cref{mainthm:equipartition-implies-reversibility-by-TO} when the min- and
max-relative entropies converge to the same value.  Whether this observation is
relevant in physically interesting systems is an open question.

It seems plausible that our results could be robust to slight violations of
translation invariance.  
For example, slight spatial inhomoginuity in a hydrodynamic mode could be allowed.
Also, ergodic states exhibit some similarities with
states obeying the eigenstate thermalization
hypothesis~\cite{Srednicki1994PRE_ETH,%
  Rigol2008Nat_ETH,DAlessio2016AP_chaos}, such as exponential decay of
off-diagonal entries of the density matrix~\cite{Sagawa-CMP-inprep}, suggesting
that our techniques could be extended to such settings.  Furthermore, a
characterization of infinite or continuous mixtures of ergodic states is
lacking, as opposed to the finite mixture considered in
\cref{mainthm:thermo-potential-iff-mix-ergodic-states-eq-potential}.  Finally,
one might hope that our methods can be extended to models exhibiting disorder,
where a gap between the min- and max-relative entropies would characterize the
irreversibility of conversions between many-body-localized states.

\begin{acknowledgments}
The authors are grateful to
Matteo Lostaglio,
Keiji Matsumoto,
Yoshiko Ogata,
and
Hiroyasu Tajima
for valuable discussions.
TS is supported by JSPS KAKENHI Grant Number JP16H02211 and JP19H05796.
PhF is supported by the Institute for Quantum Information and Matter (IQIM) at
Caltech which is a National Science Foundation (NSF) Physics Frontiers Center
(NSF Grant {PHY}-{1733907}), by the Department of Energy Award
\hbox{DE-SC}{0018407}, and by the Swiss National Science Foundation (SNSF) via
the NCCR QSIT as well as project No.~{200020\_165843}.
KK acknowledge funding provided by the Institute for Quantum Information and
Matter, an NSF Physics Frontiers Center (NSF Grant {PHY}-{1733907}).
FB is is supported by the NSF.
\end{acknowledgments}

\def\ {\unskip\space}
\def\doibase#110.{https://doi.org/10.}
\csdef{select@language}#1{}
%

\end{document}